\documentclass[twocolumn, epl, showpacs]{revtex4}
\usepackage{amssymb}
\usepackage{graphicx}
\usepackage{dcolumn}
\usepackage{bm}
\usepackage{amsmath}
\begin{document}
\title{Spontaneous trimerization in two-dimensional antiferromagnets}
\author{Zi Cai}
\affiliation{ Beijing National Laboratory for Condensed Matter
Physics, Institute of Physics, Chinese Academy of Sciences, Beijing
100080, P. R. China}
\author{Shu Chen}
\affiliation{ Beijing National Laboratory for Condensed Matter
Physics, Institute of Physics, Chinese Academy of Sciences,
Beijing 100080, P. R. China}
\author{Yupeng Wang}
\affiliation{ Beijing National Laboratory for Condensed Matter
Physics, Institute of Physics, Chinese Academy of Sciences, Beijing
100080, P. R. China}

\date{Received \today }

\begin{abstract}
In this paper, we propose a novel quantum paramagnetic state in two-dimensional antiferromagnets: the spontaneously trimer state, which is
constructed by the trimers of spins. Each trimer is a singlet state formed by three neighboring spins with SU(3) symmetry. A frustrated spin-1
Heisenberg model in the kagome lattice is investigated. By analogy to the pseudo-potential approach in the fractional quantum Hall effect
(FQHE), we find that the trimer state provides a perfect description for the exact ground state of this model. Other interesting properties,
such as the local excitations as well as magnetization plateaus have also been investigated.
\end{abstract}

\maketitle

The nature of quantum paramagnetic phases of the two-dimensional
(2D) antiferromagnetic systems attracted considerable attention in
the past twenty years because of its potential association with the
physics of the cuprate superconductors \cite{Anderson}. Because of
absence of long-range order in the quantum paramagnetic phase, a
central problem is the classification of the phases and critical
points. According to their symmetries, we could classify these
phases into two classes. The first kind is known as ``spin liquid
state", which restores the SU(2) symmetry of the Neel state and does
not break any symmetry of the original Hamiltonian. One of the
well-known state in this class is the resonating valence bond (RVB)
state \cite{Fazekas}, or the ``spin liquid" state \cite{Affleck}.

On the other hand, Read and Sachdev have presented another
possibility \cite{Read} based on the Schwinger boson analysis on the
SU(N) quantum antiferromagnets. Their key point is that the
condensation of the instantons with the Berry phase leads to the
spontaneous breaking of the spatial symmetry (translational or
rotational symmetry) of the original Hamiltonian. Similar with the
solid, this state possesses a short range order and thus is
nominated as valence-bond solids (VBS) state. As shown in
Ref.\cite{Read}, the structure of the VBS state depends on 2S(mod
4), where S is the spin of the SU(N) model. For the usual spin 1/2
systems, corresponding VBS state is known as ``dimer state", where
two nearest spins form a singlet or a dimer:
$(\uparrow\downarrow-\downarrow\uparrow)/\sqrt{2}$ and the overall
state is the product of all those dimers.

The dimer state is first proposed as the exact ground state for the
Majumdar-Ghosh(MG) model \cite{Majumdar}, which is an
antiferromagnetic spin chain with nearest and next-nearest neighbor
interactions. For a particular ratio of those exchange interactions,
the model has a two-foldly degenerate dimer ground state. Inspired
by the exact solution of the MG model, Shastry and Sutherland
proposed a two dimensional model \cite{Shastry} known as
Shastry-Sutherland (SS) model, whose ground state is the exact dimer
state in 2D. It has been used in understanding the physical
properties of $SrCu_3(BO_3)_2$ \cite{Kageyama}, which is
topologically equivalent to the SS model.

Most above researches focus on the spin-1/2 system, because of its
potential correlation with the high-Tc superconductivity. However,
the cold atom systems have provided an ideal playground to
investigate the high-spin strongly correlated system experimentally.
For example, several proposals have been provided to realize the
spin-1 Heisenberg model in the optical lattice \cite{Ripoll}. Even
in one dimension, new possibility emerges and a typical example is
known as the Affleck-Kennedy-Lieb-Tasaki (AKLT) state \cite{AKLT}.
This state breaks no symmetry of the Hamiltonian and is believed to
provide a good description for the ground state of the pure spin-1
Heisenberg chain \cite{Arovas,Cai}. As to the 2D case, the situation
is apparently more complex, the VBS state in the spin-1 quantum
paramagnetic phase could be either the dimer state \cite{Yip} that
spontaneously breaks the translational symmetry, or a generalized
AKLT state \cite{Cai} that only breaks the rotational symmetry.

In this paper, we present a novel VBS state named as ``trimer state"
for the spin-1 system in the 2D Kagome lattice. Such a state is a
product of trimer singlets with each spin singlet formed by three
spins with the total spin zero. Inspiring by the seminal work of
Ref. \cite{Arovas}, we constructed a frustrated Heisenberg model in
a kagome lattice, and found that at a particular ratio of the
different couplings, the ``trimer state" provides a perfect
approximation for the exact ground state of this model. Similar
states have been discussed in an SU(3) spin tetrahedron chain
\cite{Chen2,Greiter1,Greiter2,Solyom} and in the kagome lattice with
distorted coupling\cite{Hida1}. To our knowledge, we provide the
first example in an isotropic 2D SU(2) antiferromagnets with uniform
coupling, where the spins are spontaneously trimerized. Then we
investigated the excited properties and the magnetization plateaus
in the applied magnetic field. It is believed that this state
represents a novel class of the VBS state: N neighboring spins
cluster together to form a singlet with SU(N) symmetry and
spontaneously break the spacial symmetry. Within a cluster, each
spin is maximally entangled with other spins. We call this kind of
state as ``cluster state" while the usual dimer state is the
simplest case.
\begin{figure}[htb]
\includegraphics[width=3.2in]
{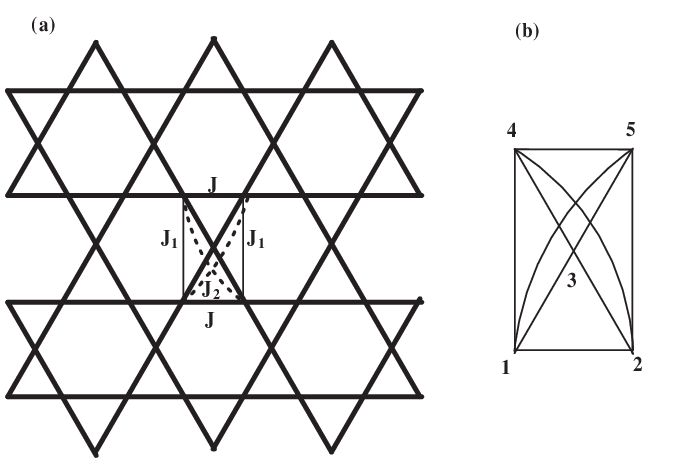} \caption{(a) The structure of frustrated Heisenberg
model in kagome lattice, with the nearest coupling $J$, the second
nearest coupling $J_1$ and the third nearest coupling $J_2$. (b)
One block consists of 5 sites and each one equally couples with
others } \label{fig1}
\end{figure}

The ground state of the quantum antiferromagnetic Heisenberg model
on the kagome lattice has been intensively investigated by various
methods, but is still far from being totally understood even for
the spin-1/2 case. Recently, Laws et. al found a $Ni^{2+}$-based
material: $Ni_3V_2O_8$ \cite{Lawes}, which is topologically
equivalent to a frustrated spin-1 Heisenberg model in kagome
lattice. The spin-1 model on the kagome lattice has been
investigated in several works \cite{Xu1,Xu2,Levin,Isakov}, but
most of them focus on the uniaxial anisotropic case. Here we focus
on the SU(2) case and introduce the second nearest $\langle\langle
\mathbf{ij}\rangle\rangle$ and the third nearest $[\mathbf{ij]}$
couplings $J_1$ and $J_2$. Explicitly, the Hamiltonian for this
model is given by
\begin{equation}
H=J\sum_{\langle \mathbf{i},\mathbf{j}\rangle
}\mathbf{S}_{\mathbf{i} }\cdot
\mathbf{S}_{\mathbf{j}}+J_1\sum_{\langle \langle
\mathbf{i},\mathbf{j} \rangle \rangle
}\mathbf{S}_{\mathbf{_i}}\cdot \mathbf{S}_{\mathbf{_j}}+J_2\sum_{[
\mathbf{i},\mathbf{j}] }\mathbf{S}_{\mathbf{i} }\cdot
\mathbf{S}_{\mathbf{j}},
\end{equation}
with the structure schematically shown in Fig.1.

In this paper we would focus on the point $J=3J_1=3J_2$ and
investigate the ground state at this point by the ``pseudopotential
approach", which was first used to get the famous Laughlin
wavefunction in the FQHE \cite{Haldane}. The analogy between the
Heisenberg model and the FQHE was first introduced by Arovas et. al.
\cite{Arovas} in the 1D case, where the Hamiltonian of a spin-1
antiferromagnetic Heisenberg model is decomposed into the summation
of projection operators
\begin{equation}
H_1=\sum_{i} \mathbf{S_i}\cdot \mathbf{S_{i+1}} = \sum_i [3
P^2(i,i+1)+P^1(i,i+1)-2].
\end{equation}
Observing that the AKLT state \cite{AKLT} is the exact ground state
of the first part of the summation of projection operators $ \sum_i
3P^2(i,i+1)$, they take it as a trial ground state of the spin-1
Heisenberg chain and consider the second part $\sum_i P^1(i,i+1)$ as
a perturbation. It turns out that the AKLT state is a very good
approximation of the exact ground state of Eq. (2) and numerical
results show that the difference between the ground state energies
of these two states is within $5\%$. Recently, we generalized this
method to the 2D spin-1 $J_1-J_2$ antiferromagnetic model\cite{Cai},
and found that at the maximal frustrated point ($J_1=2J_2$), the
ground state could be described by a two-fold 2D generalized AKLT
states, which completely agrees with the general prediction by the
field theory\cite{Read}, and was verified by the numerical results
recently\cite{Jiang}.

Next we use the similar method to investigate the ground state of
Eq. (1). Notice that at the point $J=3J_1=3J_2$, Eq. (1) could be
rewritten as the sum of identical blocks $\mathfrak{B}_\alpha$, as
shown in Fig 1.(b). Each block is constructed by 5 spins and every
spin is coupled to the other four spins identically, thus we have
\begin{equation}
 H/J_1 =\sum_\alpha \mathfrak{B}_\alpha
\quad with\quad \mathfrak{B}_\alpha =\sum_{\mathbf{{i},{j}\in \alpha
}}\mathbf{S}_{\mathbf{_i}}\cdot {\bf S}_{\mathbf{_j}}
\end{equation}
where $\alpha$ denotes the $\alpha th$ block. Then we expand this
block Hamiltonian by the projection operators of the total spin in
a block. In terms of the projection operators, we have
\begin{equation}
\mathfrak{B}_\alpha=\sum_{S=0}^5C_S\mathbf{P_\alpha ^S}-5 \quad
with\quad C_s=S(S+1)/2.
\end{equation}
To make progress, we further decompose it as two part
\[
\mathfrak{B}_\alpha =\mathfrak{B}_\alpha ^0+\mathfrak{B}_\alpha ^1,
\]
with $\mathfrak{B}_\alpha ^0=15\mathbf{P_\alpha ^5}+10\mathbf{P_\alpha ^4}+6\mathbf{P_\alpha ^3}-5$ and $%
\mathfrak{B}_\alpha ^1=3\mathbf{P_\alpha ^2}+\mathbf{P_\alpha
^1}$. The operator $ P_\alpha ^S$ projects the spin state of the
$\alpha $th Block onto the subspace with total spin $S$. Dividing
the original Hamiltonian into two parts is inspired by the success
of Haldane's pseudopotential method in FQHE \cite{Haldane} and the
model of the spin-1 chain \cite{Arovas}. Return to our spin model,
now it is clear why we divided the Hamiltonian like this: notice
that the coefficient decreases rapidly as $S$ descends, thus if we
can find the exact ground state of
$\mathfrak{B}_0=\sum_\alpha\mathfrak{B}^0_\alpha$, we can treat
the left part $\mathfrak{B}_1 =\sum_\alpha
\mathfrak{B}^{1}_\alpha$ and investigate the properties of the
ground state.
\begin{figure}[htb]
\includegraphics[width=3.2in]
{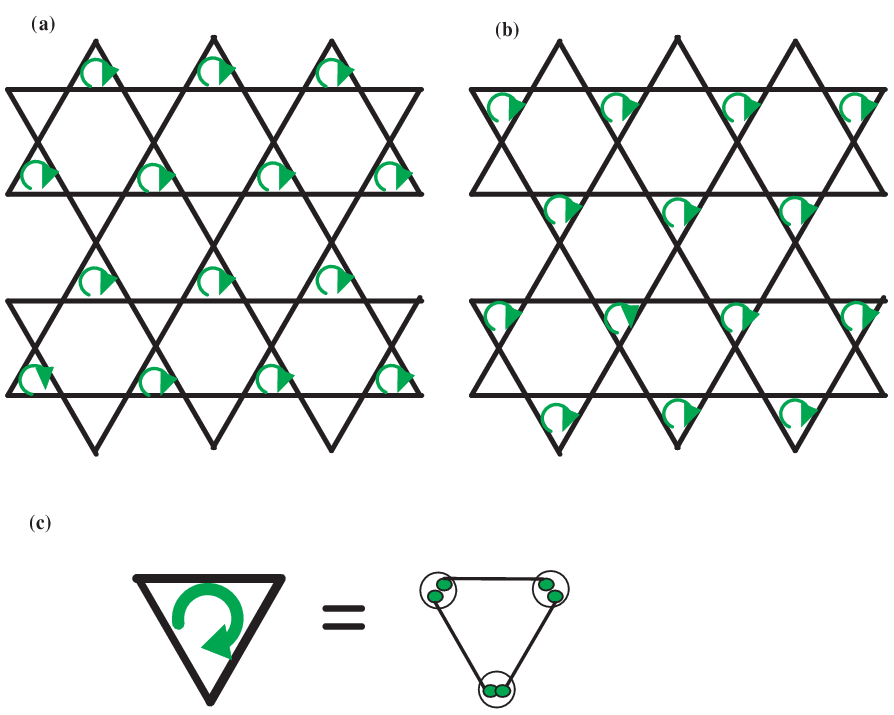} \caption{(color online)(a) and (b): The two-fold
degenerate ground states: $|\Psi_a\rangle$ and $|\Psi_b\rangle$. (c)
The structure of one trimer: a AKLT chain with Length L=3 and
periodical boundary condition} \label{fig2}
\end{figure}

First we focus on the exact ground state of the model
\begin{equation}
\mathfrak{B}_0=\sum_\alpha \left[ 15\mathbf{P_\alpha
^5}+10\mathbf{P_\alpha ^4}+6\mathbf{P_\alpha ^3}-5\right] .
\end{equation}
Since $\mathfrak{B}_\alpha ^0$ is positive semidefinite, any state
with the total spin of each block $S^T_\alpha\leq2$ is the exact
ground state of Eq.(5). It is not difficult to find that the only
possibilities are the two-fold degenerate states as shown in
Fig.2(a) and (b), where each block possesses a trimer singlet.
Explicitly, the trimer singlet can be represented as
\[
|T\rangle=\frac 1{\sqrt{6}}\sum_{\alpha, \beta, \gamma} \epsilon
_{\alpha \beta \gamma }\left| \alpha _i,\beta _j,\gamma
_k\right\rangle
\]
which is a singlet composed of three spins on neighboring sites
$i,$ $j,$ and $k$ with total spin zero (Fig.2(c). Here $\alpha _i$
denotes the spin on site $i$ with the value $\alpha $ taking $1,$
$0,$ or $-1$ and $\epsilon _{\alpha \beta \gamma }$ is an
antisymmetric tensor. Because there exists a trimer singlet in
each block, the total spin of each spin block could not be larger
than 2. Therefore it is straightforward that we have found the
exact ground states of $\mathfrak{B}_0$, or in other words, the
trial ground state of our original Hamiltonian Eq. (3).

Now we will discuss the effect of the perturbation part. Since the
trial ground states are two-fold degenerate, there seems to be a
possibility that the perturbation would resonate these two trial
ground states $|\Psi_a\rangle$ and $|\Psi_b\rangle$ to form a linear
superposition and further lower the energy of the ground state.
However, we would show this is not the case, at least in the
thermodynamic limit. Let's calculate the nondiagonal term
$\langle\Psi_a|H|\Psi_b\rangle$, it is not difficult to find that
this nondiagonal term vanishes in the thermodynamic
limit($N\rightarrow\infty$), just like the 1D MG model. So the
perturbation would not shift the two-fold degeneracy of the trial
ground states and we can safely conclude that $|\Psi_a\rangle$ and
$|\Psi_b\rangle$ provide a good description for the Hamiltonian Eq.
(3). It is straightforward to get the ground energy of this
variational ground state at the point $J=3J_1=3J_2$
\[
\langle\Psi_a|H|\Psi_a\rangle=-\frac 32JN,
\]
where N is the number of the triangles in the lattice.

It is also interesting to study the properties of the excited
states as well as magnetization process of this model. It's well
known that for the 2D dimer model, there are magnetization
plateaus in the magnetization curve because of the localization of
the single triplet excitation \cite{Kageyama,Miyahara,Knetter}. It
is natural to ask whether it happens in our model. Notice that for
a single trimer, the lowest excitation state is the triplet state,
which is the product of a dimer singlet composed of two spins
($\frac 1{\sqrt{3} }(|1,-1\rangle +|-1,1 \rangle -|00\rangle$) and
a single spin state ($|1\rangle,|-1\rangle$ or $|0\rangle$).
\begin{figure}[htb]
\includegraphics[width=3.2in]
{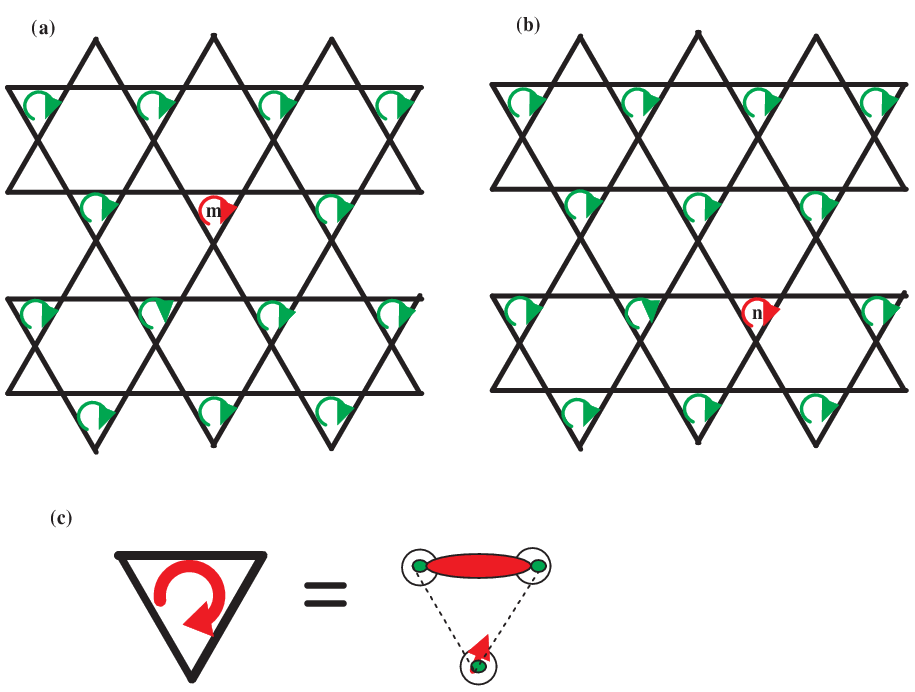} \caption{The triplet excited states: a triplet locates
at m (a) and n (b), m and n are nearest neighbor triplets. A triplet
is formed by a dimer of two spin 1 and a single spinon (c)}
\label{fig3}
\end{figure}

 The triplet state is shown in Fig.3. For convenience, we use $|m\rangle$ denoting the $m$th
 trimer singlet being excited to a triplet. To study the low energy
 excitation of this model. We calculate and find that
\begin{equation}
\langle m|H|n\rangle = 0
\end{equation}
for $m \not= n$ and
\begin{equation}
\langle m|H|m\rangle=-J-\frac 32J(N-1)
\end{equation}
where $H$ is the Hamiltonian (3), and $|m\rangle$ and $|n\rangle$
are shown in Fig.3 (a), (b). Eq.(6) means that the single triplet
excitation is almost dispersionless, at least to the order we
considered. Eq.(7) means that it is gapped, with a gap of $\frac
12J$. This dispersionless triplet excitation in our model is
different from its one dimensional analogue \cite{Takano,Nakano},
and it would localize or form a bound pair with other single
triplet, just as in the dimer state \cite{Totsuka,Momoi}. Because
of the localization of the single triplet excitation, there are
magnetization plateaus in the magnetization curve in our model.
The naive picture for the magnetization process is shown in Fig.4.
Therefore, at least two magnetization plateaus appear at
$m/m_{sat}$=1/2 and 3/4 corresponding to the phase (b) and (c) in
Fig.4. It is known that similar magnetization process and
magnetization plateaus have been studied in other Spin-1 systems,
such as S=1 spin chain\cite{Nakano}, or S=1 Heisenberg model (only
including the nearest coupling) in uniform and distorted kagome
lattice\cite{Hida}.

However, we have not considered the effect of the interactions
between the triplets, which may lead to new plateaus. Further more,
it is possible that the higher order interaction would result in the
pair of neighboring triplets, similar to the dimer phase
\cite{Totsuka,Momoi}.

\begin{figure}[htb]
\includegraphics[width=3.0in]
{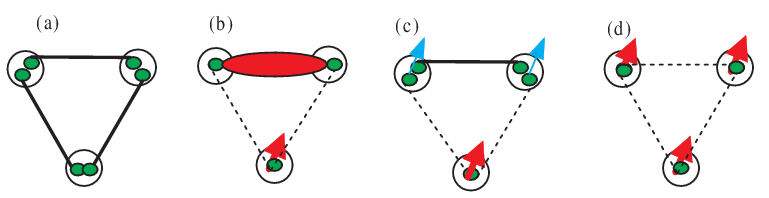} \caption{The magnetization process for a single trimer.(a)The ground state with total spin $S_{total}$=0 .(b)The state with total
spin $S_{total}$=1, it is formed by a dimer and a spinon with S=1(the bigger red arrowhead). (c)$S_{total}$=2, the smaller blue arrowhead
represents the spinon with S=1/2. (d)$S_{total}$=3, the state with saturated magnetization} \label{fig4}
\end{figure}

Now we will discuss the stability of our trimer state. Up to now, all of our analysis are based on a special point of our coupling parameters: .
At this point, the trimer state actually provide a good approximation of the exact ground state, and slight perturbation should not change the
nature of this ground state because it is a gapped VBS state which can not change by small perturbation unless it can overcome the energy gap,
just as in one dimensional AKLT state, introducing small next nearest coupling can't change the nature of 1D VBS state\cite{Nakano2} .

Usually there are two kinds of perturbation, the first kind is
deviation from the special point of the coupling parameters
$J=3J_1=3J_2$, as analyzed above, where slight deviation will not
change the nature of the ground state. However, there is another
important point: $J_1=J_2=0$, which corresponds to the S=1
antiferromagnetic Heisenberg model with only nearest coupling.
Hida first studied the ground state in this point by means of
exact diagonalization and the cluster expansion \cite{Hida1}. It
is shown that the ground state in this case is the hexagon singlet
solid (HSS) state, rather than the trimer state in our case. The
second possible perturbation for this kind of spin-1 system is the
spin biquadratic term, which is necessary to construct the 1D AKLT
model\cite{AKLT}. However, as shown in Ref.\cite{Arovas}, the
exact ground state of the AKLT Hamiltonian ($3P^2(i,i+1)$ in
Eq.(2)) which includes the biquadratic term, actually provides a
perfect variational ground state for the pure S=1 Heisenberg
Hamiltonian without biquadratic term. Therefore in our case, the
perturbation of including the small spin biquadratic term would
not change the nature of our trimer state.

\begin{figure}[htb]
\includegraphics[width=3.0in]
{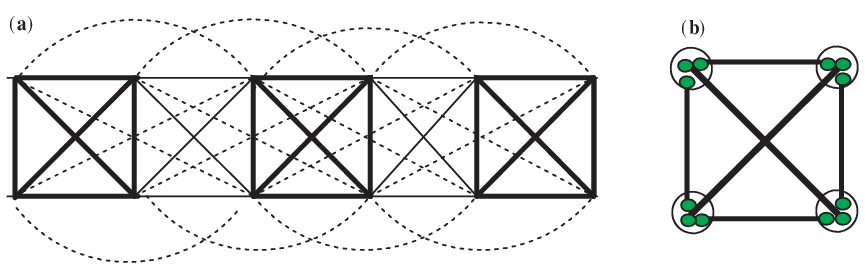} \caption{(a) The cluster state (N=4) in spin-3/2 ladder. (b)The structure of an SU(4) cluster singlet. A 3/2 spin can be
considered as a symmetrized state of three 1/2 spins. The bold lines represent a singlet for two 1/2 spins} \label{fig3}
\end{figure}

The trimer state is not the only example of our cluster state,
another example is the spin-3/2 SU(4) spin ladder \cite{Chen}. At
particular value of the coupling, the ground state of this model
is the exact spontaneous plaquette ground states, in which four
3/2 spins form an SU(4) spin singlet plaquette and spontaneously
breaking the translational symmetry, as shown in Fig.5. This model
is a generalization of the MG model to the spin ladder systems and
its ground state belongs to our cluster state with $N=4$.

After completing this work, we learnt that in a recent
paper\cite{Rico}, similar topic about the 2D multipartite VBS
state has been discussed. Z.C. thanks Flavio S. Nogueira for
helpful discussions. This work is partially supported by NSF of
China under Grant No. 10574150, MOST grant 2006CB921300 and
programs of Chinese Academy of Sciences and Deutscher Akademischer
Austausch Dienst (DAAD).

\end{document}